\documentclass[aps,prl,twocolumn,superscriptaddress,showpacs]{revtex4-1}

\usepackage{graphicx}
\usepackage{tabularx}
\usepackage{pdftexcmds}
\usepackage[T1]{fontenc}

\begin{document}


\title{Observation of Energy and Baseline Dependent Reactor Antineutrino Disappearance in the RENO Experiment}

%

\affiliation{Institute for Universe and Elementary Particles, Chonnam National University, Gwangju 61186, Korea          }
\affiliation{Department of Physics, Chung Ang University, Seoul 06974, Korea                    }
\affiliation{Department of Radiology, Dongshin University, Naju 58245, Korea                     }
\affiliation{GIST College, Gwangju Institute of Science and Technology, Gwangju 61005, Korea         }
\affiliation{Department of Physics, Gyeongsang National University, Jinju 52828, Korea         }
\affiliation{Institute for Basic Science, Daejeon 34047, Korea     }
\affiliation{Department of Physics, Kyungpook National University, Daegu 41566, Korea          }
\affiliation{Department of Physics and Astronomy, Sejong University, Seoul 05006, Korea     }
\affiliation{Department of Physics and Astronomy, Seoul National University, Seoul 08826, Korea }
\affiliation{Department of Fire Safety, Seoyeong University, Gwangju 61268, Korea              }
\affiliation{Department of Physics, Sungkyunkwan University, Suwon 16419, Korea                }

\author{J. H. Choi}
\affiliation{Department of Radiology, Dongshin University, Naju 58245, Korea                     }
\author{W. Q. Choi}
\affiliation{Department of Physics and Astronomy, Seoul National University, Seoul 08826, Korea }
\author{Y. Choi}
\affiliation{Department of Physics, Sungkyunkwan University, Suwon 16419, Korea                }
\author{H. I. Jang}
\affiliation{Department of Fire Safety, Seoyeong University, Gwangju 61268, Korea              }
\author{J. S. Jang}
\affiliation{GIST College, Gwangju Institute of Science and Technology, Gwangju 61005, Korea         }
\author{E. J. Jeon}
\affiliation{Institute for Basic Science, Daejeon 34047, Korea     }
\affiliation{Department of Physics and Astronomy, Sejong University, Seoul 05006, Korea     }
\author{K. K. Joo}
\affiliation{Institute for Universe and Elementary Particles, Chonnam National University, Gwangju 61186, Korea          }
\author{B. R. Kim}
\affiliation{Institute for Universe and Elementary Particles, Chonnam National University, Gwangju 61186, Korea          }
\author{H. S. Kim}
\affiliation{Department of Physics and Astronomy, Sejong University, Seoul 05006, Korea           }
\author{J. Y. Kim}
\affiliation{Institute for Universe and Elementary Particles, Chonnam National University, Gwangju 61186, Korea          }
\author{S. B. Kim}
\affiliation{Department of Physics and Astronomy, Seoul National University, Seoul 08826, Korea }
 \author{S. Y. Kim}
\affiliation{Department of Physics and Astronomy, Seoul National University, Seoul 08826, Korea }
\author{W. Kim}
\affiliation{Department of Physics, Kyungpook National University, Daegu 41566, Korea          }
\author{Y. D. Kim}
\affiliation{Institute for Basic Science, Daejeon 34047, Korea     }
\affiliation{Department of Physics and Astronomy, Sejong University, Seoul 05006, Korea                       }
\author{Y. Ko}
\affiliation{Department of Physics, Chung Ang University, Seoul 06974, Korea                    }
\author{D. H. Lee}
\affiliation{Department of Physics and Astronomy, Seoul National University, Seoul 08826, Korea }
\author{I. T. Lim}
\affiliation{Institute for Universe and Elementary Particles, Chonnam National University, Gwangju 61186, Korea          }
\author{M. Y. Pac}
\affiliation{Department of Radiology, Dongshin University, Naju 58245, Korea                     }
\author{I. G. Park}
\affiliation{Department of Physics, Gyeongsang National University, Jinju 52828, Korea         }
\author{J. S. Park}
\affiliation{Department of Physics and Astronomy, Seoul National University, Seoul 08826, Korea }
\author{R. G. Park}
\affiliation{Institute for Universe and Elementary Particles, Chonnam National University, Gwangju 61186, Korea          }
\author{H. Seo}
\affiliation{Department of Physics and Astronomy, Seoul National University, Seoul 08826, Korea }
\author{S. H. Seo}
\affiliation{Department of Physics and Astronomy, Seoul National University, Seoul 08826, Korea }
\author{Y. G. Seon}
\affiliation{Department of Physics, Kyungpook National University, Daegu 41566, Korea          }
\author{C. D. Shin}
\affiliation{Institute for Universe and Elementary Particles, Chonnam National University, Gwangju 61186, Korea          }
\author{K. Siyeon}
\affiliation{Department of Physics, Chung Ang University, Seoul 06974, Korea                    }
 \author{J. H. Yang}
\affiliation{Department of Physics, Sungkyunkwan University, Suwon 16419, Korea                }
\author{I. S. Yeo}
\affiliation{Institute for Universe and Elementary Particles, Chonnam National University, Gwangju 61186, Korea          }
\author{I. Yu}
\affiliation{Department of Physics, Sungkyunkwan University, Suwon 16419, Korea                }

\collaboration{The RENO Collaboration}

%
%

\begin{abstract}

The RENO experiment has analyzed about 500 live days of data to observe an energy dependent disappearance of reactor $\overline{\nu}_e$ by comparison of their prompt signal spectra measured in two identical near and far detectors. In the period between August 2011 and January 2013, the far (near) detector observed 31541 (290775) electron antineutrino candidate events with a background fraction of 4.9\% (2.8\%). The measured prompt spectra show an excess of reactor $\overline{\nu}_e$ around 5 MeV relative to the prediction from a most commonly used model. A clear energy and baseline dependent disappearance of reactor $\overline{\nu}_e$ is observed in the deficit of the observed number of $\overline{\nu}_e$. Based on the measured far-to-near ratio of prompt spectra, we obtain $\sin^2 2 \theta_{13} = 0.082 \pm 0.009({\rm stat.}) \pm 0.006({\rm syst.})$ and $|\Delta m_{ee}^2| =[2.62_{-0.23}^{+0.21}({\rm stat.}) _{-0.13}^{+0.12}({\rm syst.})]\times 10^{-3}$~eV$^2$.
\end{abstract}

\pacs{14.60.Pq, 29.40.Mc, 28.50.Hw, 13.15.+g}
\keywords{neutrino oscillation, neutrino mixing angle, reactor antineutrino }

\maketitle

%

The reactor $\overline{\nu}_e$ disappearance has been firmly observed to determine the smallest neutrino mixing angle $\theta_{13}$ \cite{RENO, DB, DC}. All of the three mixing angles in the Pontecorvo-Maki-Nakagawa-Sakata matrix \cite{Ponte, MNS} have been measured to provide a comprehensive picture of neutrino transformation. The successful measurement of a rather large $\theta_{13}$ value opens the possibility of searching for CP violation in the leptonic sector and determining the neutrino mass ordering. Appearance of $\nu_e$ from an accelerator $\nu_{\mu}$ beam is also observed by the T2K~\cite{T2K} and NO$\nu$A~\cite{NOvA} experiments.

Using the $\overline{\nu}_e$ survival probability $P$ \cite{Petcov}, reactor experiments with a baseline distance of $\sim$1 km can determine the mixing angle $\theta_{13}$ and an effective squared-mass-difference $\Delta m_{ee}^2 \equiv \cos^2 \theta_{12}\Delta m_{31}^2 + \sin^2 \theta_{12} \Delta m_{32}^2$ \cite{Parke}. 
\begin{eqnarray}
 1- P & =  &  \sin^2 2 \theta_{13} ( \cos^2 \theta_{12} \sin^2 \Delta_{31} + \sin^2 \theta_{12} \sin^2 \Delta_{32} )
  \nonumber      \\
 &  &    + \cos^4 \theta_{13} \sin^2 2\theta_{12} \sin^2 \Delta_{21}
  \nonumber       \\
 &  &  \hspace*{-1cm} ~ \approx   \sin^2 2 \theta_{13} \sin^2 \Delta_{ee}  + \cos^4 \theta_{13} \sin^2 2\theta_{12} \sin^2 \Delta_{21}
,~\label{eqn:Papprx}
\end{eqnarray}
where $\Delta_{ij} \equiv 1.267 \Delta m_{ij}^2 L/E$, $E$ is the $\overline{\nu}_e$ energy in MeV, and $L$ is the distance  between the reactor and detector in meters.

The first measurement of $\theta_{13}$ by RENO was based on the rate-only analysis of deficit found in $\sim$220 live days of data \cite{RENO}. The oscillation frequency $|\Delta m_{ee}^2|$ in the measurement was approximated by the measured value $|\Delta m_{31}^2|$ assuming the normal ordering in the $\nu_{\mu}$ disappearance \cite{MINOS}. In this Letter, we present a more precisely measured value of $\theta_{13}$ and our first determination of $|\Delta m_{ee}^2|$, based on the rate, spectral and baseline information (rate+spectrum analysis) of reactor $\overline{\nu}_e$ disappearance using $\sim$500 live days of data. The Daya Bay collaboration has also reported spectral measurements \cite{DB-spect}.

The RENO uses identical near and far $\overline{\nu}_e$ detectors located at 294~m and 1383~m, respectively, from the center of six reactor cores of the Hanbit (known as Yonggwang) Nulcear Power Plant.  The far (near) detector is under a 450~m (120~m) of water equivalent overburden. 
Six pressurized water reactors, each with maximum thermal output of 2.8~GW$_{th}$,
are situated in a linear array spanning 1.3 km with equal spacings. The reactor flux-weighted baseline is 410.6~m for the near detector and 1445.7~m for the far detector. 

The reactor $\overline{\nu}_e$ is detected through the inverse beta decay (IBD) interaction, $\overline{\nu}_e + p \rightarrow e^+  + n$, with free protons in hydrocarbon liquid scintillator (LS) with 0.1\% Gadolinium (Gd) as a target. The coincidence of a prompt positron signal and a mean time of $\sim$28 $\mu$s delayed signal from neutron capture by  Gd (n-Gd) provides the distinctive IBD signature against backgrounds. The prompt signal releases energy of 1.02 MeV as two $\gamma$-rays from the positron annihilation in addition to the positron kinetic energy. The delayed signal produces several $\gamma$-rays with the total energy of $\sim$8 MeV. 
The RENO LS is made of linear alkylbenzene (LAB) with fluors. A Gd-carboxylate complex was developed for the best Gd loading efficiency into LS and its long term stability \cite{RENO-GdLS}.  

Each RENO detector consists of a main inner detector (ID) and an outer veto detector (OD). The ID is contained in a cylindrical stainless steel vessel that houses two nested cylindrical acrylic vessels \cite{RENO-acrylic}. The innermost acrylic vessel holds 16 tons of Gd-doped LS as a neutrino target, and is surrounded by a $\gamma$-catcher region with a 60 cm thick layer of undoped LS inside an outer acrylic vessel. Outside the $\gamma$-catcher is a 70~cm thick buffer region filled with mineral oil. Light signals emitted from particles are detected by 354 low background 10-inch photomultiplier tubes (PMTs) \cite{RENO-PMT} that are mounted on the inner wall of the stainless steel container. The 1.5~m thick OD region is filled with highly purified water, and equipped with 67 10-inch PMTs mounted on the wall of the concrete OD vessel.

Event triggers are based on the number of hit PMTs with signals above a $\sim$0.3 photoelectron (p.e.) threshold (NHIT). An event passes trigger selection and is recorded if the ID NHIT is larger than 90, corresponding to 0.5$-$0.6~MeV and well below the 1.02~MeV minimum energy of an IBD positron signal. The event energy is determined from the total charge ($Q_{tot}$) in p.e. that is collected by the PMTs within $-$100~ns to $+$50~ns and corrected for gain and charge collection variations using the neutron capture peak energies. 

An accurate energy measurement is essential for extracting $|\Delta m_{ee}^2|$ and $\theta_{13}$ from the spectral distortion of IBD prompt events that is developed by neutrino oscillation.
An absolute energy scale is determined by $Q_{tot}$ of $\gamma$-rays coming from radioactive sources of $^{137}$Cs, $^{68}$Ge, $^{60}$Co, $^{252}$Cf, and $^{209}$Po-Be, and from IBD delayed signals of neutron capture on Gd. A charge-to-energy conversion function is generated from the peak energies of these $\gamma$-ray sources. 
The observed $Q_{tot}$ of a $\gamma$-ray source is converted to the corresponding $Q_{tot}$ of a positron ($Q_{tot}^c$) using a \textsc{geant4} Monte Carlo simulation (MC). 
The true energy ($E_{true}$) of a positron interaction is the sum of the kinetic energy and the energy from its annihilation.
The converted $Q_{tot}^c$ of IBD prompt energy ($E_p$) is estimated by taking into account difference in the visible energies of $\gamma$-ray and positron through the MC. 
The RENO MC includes measured optical properties of LS and quenching effect of $\gamma$-ray at low energies \cite{RENO-GdLS}.
The quenching effect depends on the energy and the multiplicity of $\gamma$-rays released from the calibration sources. 
The MC simulated $Q_{tot}$ well reproduces that of a $\gamma$-ray source including the quenching effect depending on the multiplicity of $\gamma$-rays.  
The measured $Q_{tot}^c$ shows non-linear response to $E_{true}$, especially at low energies, mainly due to quenching effect in the scintillator and Cherenkov radiation.

\begin{figure}[hbt]
\begin{center}
\includegraphics[width=0.47\textwidth]{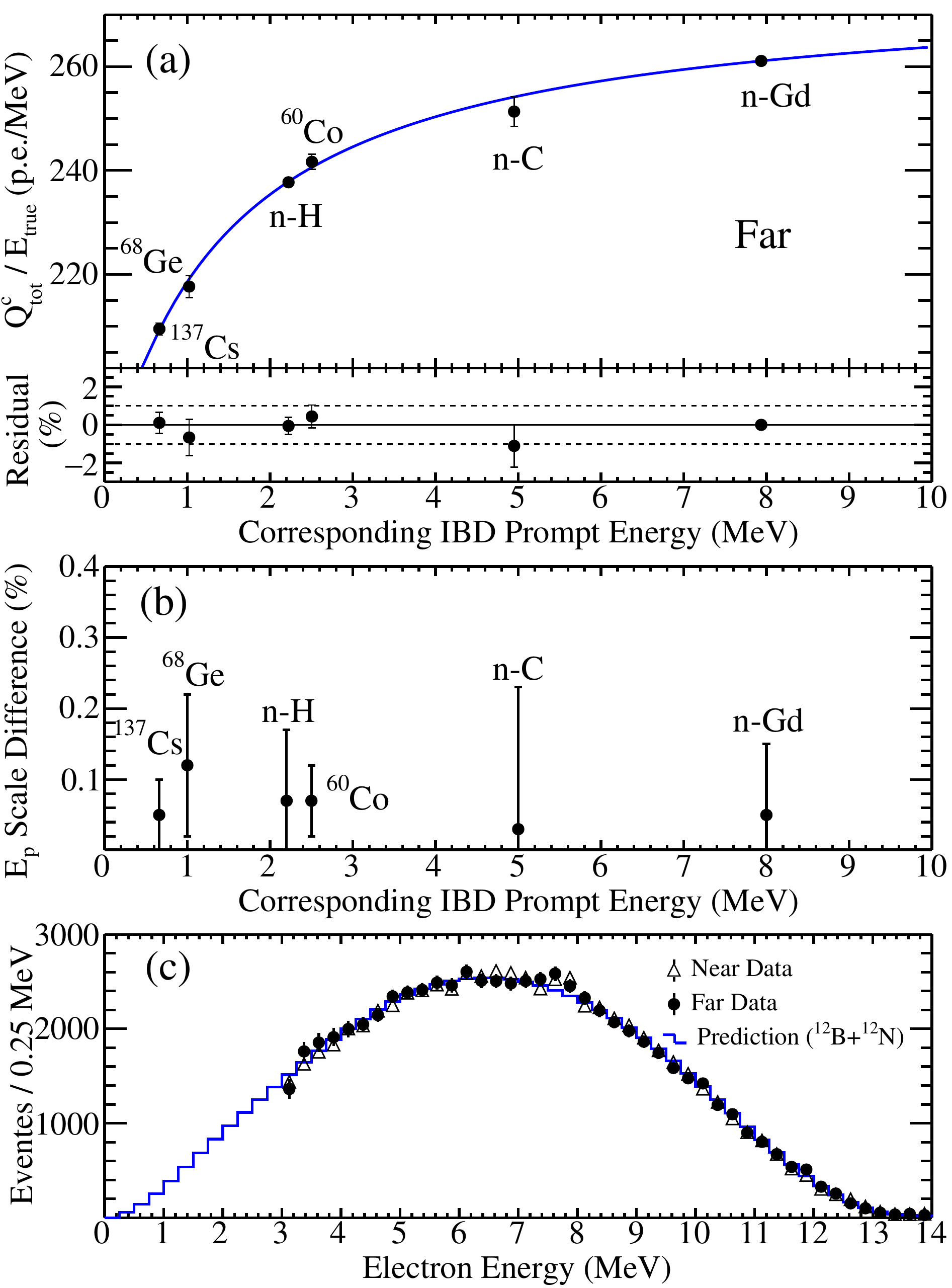}
\caption{(a) Non-linear response of scintillating energy obtained from the visible energies of $\gamma$-rays coming from several radioactive sources and IBD delayed signals in the far detector. The curve is the best fit to the data points. Note that the n-C sample is obtained from the $^{209}$Po-Be source. (b) Energy scale difference of the near and far detectors. A correlated energy scale uncertainty does not contribute to the difference due to the identical near and far detectors.
(c) Comparison of measured and simulated energy spectra of electron from $\beta$-decay of unstable isotope $^{12}$B, with minute contribution from $^{12}$N, produced by cosmic muons. }
\label{fig:energy-calibration}
\end{center}
\end{figure}

Figure \ref{fig:energy-calibration} (a) upper panel shows non-linear response of scintillating energy for the IBD prompt signal which is well described by a parametrization of $Q_{tot}^c/E_{true} = a + b/[1- \exp(-c E_{true}+d)]$. The parameters $a$, $b$, $c$, and $d$ are determined by a fit.
Deviation of all calibration data points with respect to the best-fit is within 1\% as shown in Fig. \ref{fig:energy-calibration} (a) lower panel. 
The energy scales of the near and far detectors are compared using identical radioactive sources, and the difference is found to be less than 0.15\% for $E_p = 1$$-$8~MeV as shown in Fig. \ref{fig:energy-calibration} (b). 
Figure \ref{fig:energy-calibration} (c) shows an excellent agreement between data and MC as well as between the near and far data in the electron energy spectrum of $\beta$-decays from radioactive isotopes $^{12}$B and $^{12}$N that are produced by cosmic-muon interactions. This demonstrates that the obtained parametrization for non-linear response of electron scintillating energy works well for energies of 3 to 14~MeV within the statistical fluctuation of the data sample.  

Event selection criteria are applied to obtain clean IBD candidates with a delayed signal of neutron capture by Gd. The details are given in Ref. \cite{RENO} and added  or modified as follows: (i) extended timing veto criteria to reject events associated with muon if they are within a 700 ms (500 ms, 200 ms)  window following a cosmic muon of $E_{\mu} > 1.5$ GeV (1.2$-$1.5~GeV, 1.0$-$1.2~GeV) for the far detector and a similar set of criteria for the near detector; (ii) relaxed $Q_{max}/Q_{tot}$ requirement from $< 0.03$ to $< 0.07$ to minimize possible signal loss at low energies where $Q_{max}$ is the maximum charge of any single ID PMTs;
(iii) $\Delta R < 2.5$~m for additional reduction of accidental backgrounds, where $\Delta R$ is the distance between the prompt and delayed signals; (iv) additional PMT hit timing and charge requirements to eliminate events coming from flashing PMTs effectively if they satisfy $Q_{max}/Q_{tot} > 0.07$ where an extended timing window of -400~ns to +800 ns is imposed to calculate $Q_{tot}$ and $Q_{max}$ for this criterion; (v) multiplicity requirements for rejecting coincidence pairs if there are other pairs within 500~$\mu$s interval, or if any ID triggers other than those associated with the delayed signal candidate occurring within 200~$\mu$s from its prompt signal candidate.
The total signal loss due to the additional criteria is 11.0\% (11.4\%) with an uncertainty of 0.02\% (0.01\%) for the far (near) detector. Thus the uncorrelated systematic uncertainty of detection efficiency between the near and far detectors is hardly affected and remains to be 0.2\%. The background rate is reduced by 25.9\% (19.4\%) for the far (near) detector, with respect to the first measurement \cite{RENO}. The background uncertainty is significantly reduced from 17.7\% (27.3\%) to 7.3\% (4.7\%) for the far (near) detector.

Applying the IBD selection criteria yields 31541 (290775) candidate events with $E_p$ between 1.2 and 8.0~MeV for a live time of 489.93 (458.49) days in the far (near) detector, in the period between August 2011 and January 2013. 
IBD events with $E_p < 1.2$~MeV include prompt signals of positrons occurring in or near the target acrylic vessel that deposit kinetic energy in the acrylic without producing scintillation lights. These events are reconstructed to have visible energy near the positron annihilation energy of 1.02~MeV and are not well reproduced by the MC prediction.
The IBD signal loss by  $E_p < 1.2$~MeV requirement is roughly 2\% in both detectors.
In the final data samples, the remaining backgrounds are either uncorrelated or correlated IBD candidates. An accidental background comes from an uncorrelated pair of prompt- and delayed-like events. Correlated backgrounds are fast neutrons from outside of ID, stopping muon followers, $\beta$-$n$ emitters from cosmic muon induced $^9$Li/$^8$He isotopes, and $^{252}$Cf contamination. The total background fraction is $4.9\pm0.4$\% in the far detector, and $2.8\pm0.1$\% in the near detector.
\begin{table}[hbt]
 \caption{Observed IBD and estimated background rates at $1.2 < E_p < 8.0$~MeV given in per day.}
 \begin{center}
 \begin{tabular*}{0.48\textwidth}{@{\extracolsep{\fill}} l r r }
 
 \hline \hline
     Detector              & Near       &  Far      \\
  \hline
IBD rate                     &   $616.67\pm1.44$ &  $61.24\pm0.42$ \\
after background subtraction   &                  &                           \\
Total background rate  &   $17.54\pm0.83$ &  $3.14\pm0.23$ \\
DAQ Live time (days)    &  458.49         &  489.93   \\
    \hline
  Accidental rate    &    $6.89\pm0.09$   &  $0.97\pm0.03$  \\
  $^9$Li/$^8$He rate &    $8.36\pm0.82$   &  $1.54\pm0.23$   \\
  Fast neutron rate  &    $2.28\pm0.04$   &  $0.48\pm0.02$   \\
  $^{252}$Cf contamination rate &   $0.00\pm0.01$   &  $0.14\pm0.03$   \\ 
  \hline \hline
  \end{tabular*}
 \end{center}

 \label{tab:Event_rate}

 \end{table}  

The remaining accidental background in the final sample is estimated by measuring random spatial associations of prompt- and delayed-like events. The prompt energy spectrum of accidental background is obtained from a control sample that is selected by a requirement of temporal association larger than 1~ms. 
Even though the accidental background is increased by 42.6\% (60.2\%) in the far (near) detector, the background uncertainty remains almost the same due to high statistics of the background control sample.   
The energy spectrum of $^9$Li/$^8$He background is measured using a sample of IBD-like pairs that are preceded within 500~ms (400~ms) by energetic muons of $E_{\mu} >$ 1.5~GeV ($>$1.6~GeV) for the far (near) detector.
The $^9$Li/$^8$He background rate in the final sample is obtained from the measured rate in the background dominant region of $E_p > 8$~MeV using the measured background spectrum. The new method of $^9$Li/$^8$He background estimation contributes to reduction of the largest background uncertainty. 
The fast neutron background rate in the IBD candidates is estimated by extrapolating from the background dominant energy region, assuming a flat spectrum of the background. A fast neutron enriched sample is obtained by selecting IBD candidates if they are accompanied by any prompt candidates of  $E_p >$0.7~MeV within a 1 ms subsequent window. The prompt events of this sample show a reasonably flat spectrum in the IBD signal region. The background uncertainty includes a possible deviation from the flat spectrum, 1.3\% (1.2\%) of the fast neutron background rate for the far (near) detector.
The background rate is reduced by 50.5\% (54.4\%) due to the additional multiplicity requirements for the far (near) detector.    

A tiny amount of $^{252}$Cf was accidentally introduced into both detectors during detector calibration in October 2012. Most of multiple neutron events coming from the $^{252}$Cf contamination are eliminated by stringent multiplicity requirements.
IBD candidates are rejected: (i) if they are accompanied by any prompt candidates of  $E_p >$0.7~MeV within a 300~$\mu$s preceding window or a 1 ms subsequent window; (ii) if they are accompanied by a prompt candidate of $E_p > 3$~MeV within a 10~s window and a distance of 40~cm; (iii) if any ID and OD trigger occurs in a 200~$\mu$s window following their prompt candidates. After applying the requirements, 99.9\% of the $^{252}$Cf contamination background events in the far detector are eliminated with a signal loss of $8.0\pm0.2$\%. No remaining $^{252}$Cf contamination background events are observed in the near detector.

The total background rates are estimated to be $17.54\pm0.83$ and $3.14\pm0.23$ events per day for near and far detectors, respectively. The observed IBD and background rates are summarized in Table \ref{tab:Event_rate}. 
Since the rates and shapes of all the backgrounds are measured from control data samples, their uncertainties are expected to be further reduced with more data.

Systematic uncertainties have been significantly reduced since the first measurement presented in Ref. \cite{RENO}. Decrease of systematic uncertainties mainly comes from background reduction and more precise estimation of background rates. For example, the most dominant background uncertainty of $^9$Li/$^8$He is reduced from 29\% (48\%) to 15\% (10\%) in the far (near) detector.
The reduction was possible due to additional background removal by optimized rejection criteria, increased statistics of the $^9$Li/$^8$He control sample, and a new method of estimating the background rate in the IBD candidates from the background dominant energy region. The IBD selection criterion (i) described earlier removes 55.9\% (43.8\%) of remaining $^9$Li/$^8$He backgrounds with a signal loss of 9.7\% (10.3\%) in the far (near) detector. The uncertainty of the background spectrum is reduced because of increased control sample by a factor of five. 

The expected rate and spectrum of reactor $\overline{\nu}_e$ are calculated based on thermal power, fission fraction, energy released per fission, $\overline{\nu}_e$ yield per fission, fission spectra, and IBD cross sections \cite{Vogel, Feilitzsch, Hahn, Declais, Mueller, Huber, Kopeikin}. The calculation includes both the rate and spectral changes corresponding to the varying thermal powers and fission fractions of each reactor during data-taking. 

The systematic uncertainties in the reactor $\overline{\nu}_e$ detection are found in Ref. \cite{RENO}. The energy dependent systematic uncertainties, coming from background shape ambiguities and the energy scale difference between the near and far detectors, are evaluated and included for this analysis.

We observe a clear deficit of reactor $\overline{\nu}_e$ in the far detector. 
Using the deficit information only, a rate-only analysis obtains $\sin^2 2 \theta_{13}$ = 0.087 $\pm$ 0.009(stat.) $\pm$ 0.007(syst.), where the world average value of $|\Delta m_{ee}^2|= (2.49 \pm0.06) \times 10^{-3}$~eV$^2$ is used \cite{PDG}.
The total systematic error of $\sin^2 2 \theta_{13}$ is reduced from 0.019 to 0.007, mostly due to the decreased background uncertainty, relative to the first measurement~\cite{RENO} while the statistical error is reduced from 0.013 to 0.009.

\begin{figure}[hbt]
\begin{center}
\includegraphics[width=0.47\textwidth]{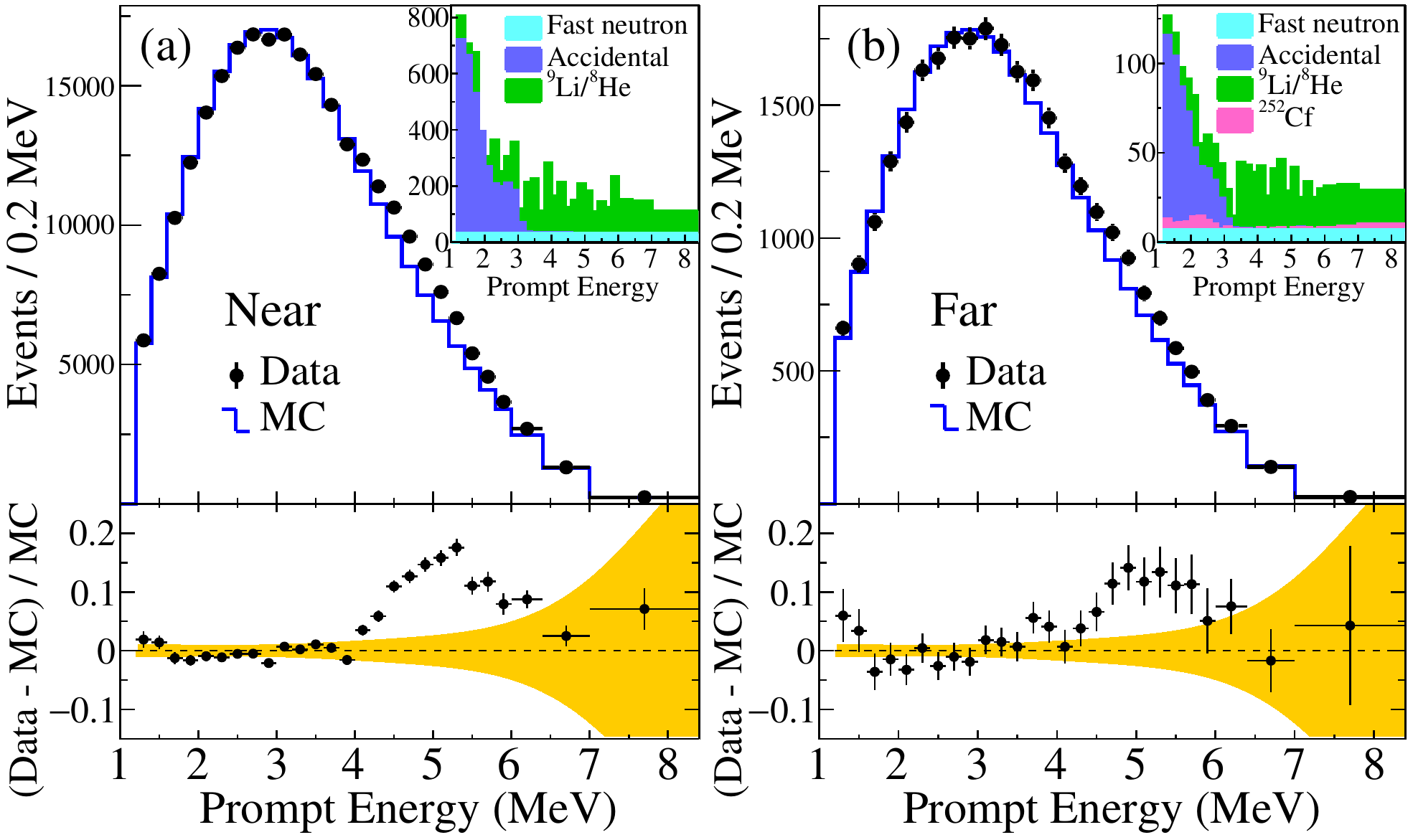}
\caption{Spectral comparison of observed and expected IBD prompt events in the (a) near and (b) far detectors. The expected distributions are obtained using rate and spectral analysis results discussed later.
The observed spectra are obtained from subtracting the background spectra as shown in the insets.
A shape difference is clearly seen at 5~MeV. The observed excess is correlated with the reactor power, and corresponds to 3\% of the total observed reactor $\overline{\nu}_e$ flux. A spectral deviation from the expectation is larger than the uncertainty of an expected spectrum (shaded band).}
\label{fig:new-reactor-neutrinos}
\end{center}
\end{figure}

Figure \ref{fig:new-reactor-neutrinos} shows a spectral comparison of the observed IBD prompt spectrum after background subtraction to the prediction that is expected from a reactor neutrino model \cite{Mueller, Huber} and the best fit oscillation results. 
The subtracted background spectra are shown in the insets.
A clear spectral difference is observed in the region centered at 5~MeV. The MC predicted distributions are normalized to the observed events out of the excess range $3.6 < E_p < 6.6$~MeV. The excess of events constitutes about 3\% of the total observed reactor $\overline{\nu}_e$ rate in both detectors. Furthermore, the excess is observed to be proportional to the reactor power. This observation suggests needs for reevaluation and modification of the current reactor $\overline{\nu}_e$ model \cite{Mueller, Huber}. 

Because of the unexpected structure around 5 MeV, the oscillation amplitude and frequency are determined from a fit to the measured far-to-near ratio of IBD prompt spectra. The relative measurement using identical near and far detectors makes the method insensitive to the correlated uncertainties of expected reactor $\overline{\nu}_e$ flux and spectrum as well as detection efficiency.
To determine $|\Delta m_{ee}^2|$ and $\theta_{13}$ simultaneously, a $\chi^2$ is constructed using the spectral ratio measurement and is minimized \cite{Anderson}:
\begin{eqnarray}
 \chi^2  & = & \sum_{i=1}^{N_{bins}} \frac{(O_i^{F/N} - T_i^{F/N})^2 }{U_i^{F/N}}
                          + \sum_{d=N, F} \left( \frac{b^{d}}{\sigma_{bkg}^{d}} \right)^2 
 \nonumber       \\
&&                        + \sum_{r=1}^{6} \left( \frac{f_r}{\sigma_{flux}^r} \right)^2 
                          + \left( \frac{\epsilon}{\sigma_{eff}} \right)^2
                          + \left( \frac{e}{\sigma_{scale}} \right)^2 ,
\end{eqnarray} 
where $O_i^{F/N}$ is the observed far-to-near ratio of IBD candidates in the $i$-th $E_p$ bin after background subtraction, $T_i^{F/N} = T_i^{F/N} (b^d, f_r, \epsilon, e; \theta_{13}, |\Delta m_{ee}^2|) $ is the expected far-to-near ratio of IBD events, and $U_i^{F/N}$ is the statistical uncertainty of $O_i^{F/N}$. 
The expected ratio $T_i^{F/N}$ is calculated using the reactor $\overline{\nu}_e$ spectrum model and the IBD cross section and folding the $\overline{\nu}_e$ survival probability and the detector effects.
The systematic uncertainty sources are embedded by pull parameters ($b^d$, $f_r$, $\epsilon$, and $e$) with associated systematic uncertainties ($\sigma_{bkg}^d$, $\sigma_{flux}^r$, $\sigma_{eff}$, and $\sigma_{scale}$ ). 
The pull parameters $b^d$ and $e$ introduce deviations from the expected IBD spectra accounting for the effects of the associated energy-dependent systematic uncertainties.
The uncorrelated reactor-flux uncertainty $\sigma_{flux}^r$ is 0.9\%, the uncorrelated detection uncertainty $\sigma_{eff}$ is 0.2\%, the uncorrelated energy scale uncertainty $\sigma_{scale}$ is 0.15\%, and the background uncertainty $\sigma_{bkg}^d$ is 4.7\% and 7.3\% for near and far detectors, respectively.
The $\chi^2$ is minimized with respect to the pull parameters and the oscillation parameters.
\begin{table}[hbt]
 \caption{Systematic errors from uncertainty sources}
 \begin{center}
 \begin{tabular*}{0.48\textwidth}{@{\extracolsep{\fill}} l c c }
  \hline \hline
                                 & $\delta |\Delta m_{ee}^2|$ ($\times 10^{-3}$ eV$^2$)    
                                 & $\delta (\sin^2 2\theta_{13})$       \\
  \hline
 Reactor                     &  $+0.018$, $-0.018$       &   $+0.0026$, $-0.0028$   \\
Detection efficiency      &  $+0.020$, $-0.022$     &   $+0.0028$, $-0.0029$  \\
Energy scale               &  $+0.081$, $-0.094$  &     $+0.0026$, $-0.0015$ \\
Backgrounds                &   $+0.084$, $-0.106$    &   $+0.0030$, $-0.0028$    \\
    \hline
 Total                        &   $+0.115$, $-0.133$     &  $+0.0055$, $-0.0052$  \\
  \hline \hline
  \end{tabular*}
 \end{center}

 \label{tab:syst_error}

 \end{table}

The best-fit values obtained from the rate and spectral analysis are $\sin^2 2\theta_{13} = 0.082 \pm 0.009(\rm stat.) \pm 0.006(\rm syst.)$ and $|\Delta m_{ee}^2| = [2.62_{-0.23}^{+0.21}({\rm stat.}) _{-0.13}^{+0.12}({\rm syst.})]\times 10^{-3}$~eV$^2$ with $\chi^2 /NDF = 58.9/66$. 
A fit result is also obtained using an independent pull parameter for each energy bin to allow maximum variation of the background shapes within their uncertainties.
The total systematic errors for both $\sin^2 2\theta_{13}$ and $|\Delta m_{ee}^2|$ remain almost unchanged.
The dominant systematic uncertainties are those of the energy scale difference and the backgrounds as shown in Table \ref{tab:syst_error}.
The measured value of $|\Delta m_{ee}^2|$ corresponds to $|\Delta m_{31}^2| = (2.64 ^{+0.24}_{-0.26})\times 10^{-3}$~eV$^2$ $( (2.60^{+0.24}_{-0.26})\times 10^{-3}$~eV$^2$) for the normal (inverted) neutrino mass ordering, using measured oscillation parameters of $\sin^2 2\theta_{12} = 0.846 \pm 0.021$ and $\Delta m_{21}^2 = (7.53 \pm 0.18)\times 10^{-5}$~eV$^2$ \cite{PDG}.
The spectral-only analysis with a free normalization yields $\sin^2 2\theta_{13} = 0.066^{+0.042}_{-0.046} $ and $|\Delta m_{ee}^2| = (2.62^{+0.38}_{-0.41}) \times 10^{-3}$~eV$^2$ with $\chi^2 /NDF = 58.8/67$.    

\begin{figure}[hbt]
\begin{center}
\includegraphics[width=0.40\textwidth]{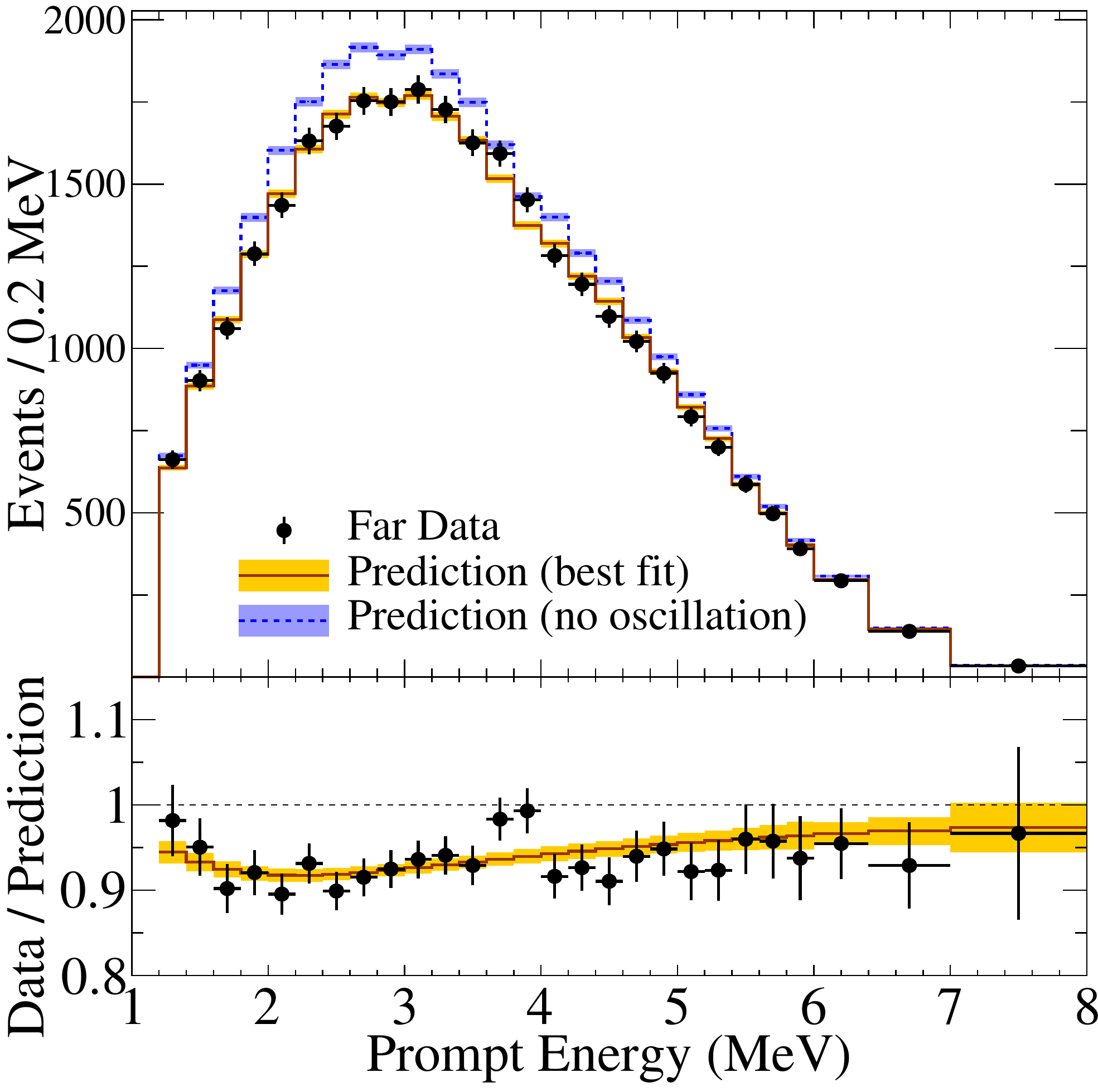}
\caption{Top: comparison of the observed IBD prompt spectrum in the far detector with the no-oscillation prediction obtained from the measurement in the near detector. The prediction from the best-fit results to oscillation is also shown. Bottom: ratio of reactor $\overline{\nu}_e$ events measured in the far detector to the no-oscillation prediction (points) and ratio from MC with best-fit results folded in (shaded band). Errors are statistical uncertainties only.}
\label{fig:spectra-oscillation-fit}
\end{center}
\end{figure}

\begin{figure}[hbt]
\begin{center}
\includegraphics[width=0.47\textwidth]{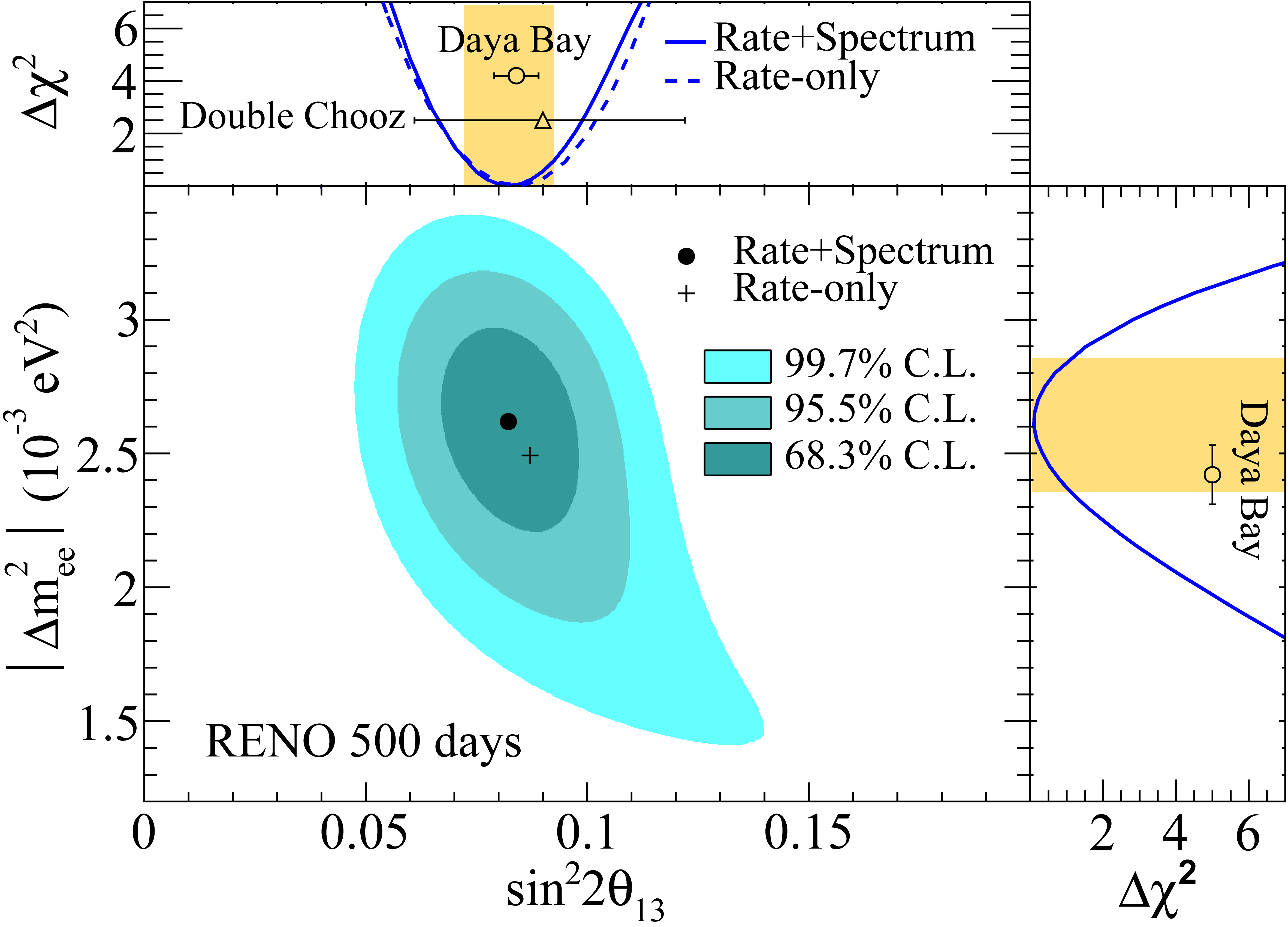}
\caption{Allowed regions of  68.3, 95.5, and 99.7\% C.L. in the $|\Delta m_{ee}^2|$ vs. $\sin^2 2\theta_{13}$ plane. The best-fit values are given by the black dot. The $\Delta \chi^2$ distributions for $\sin^2 2\theta_{13}$ (top) and $|\Delta m_{ee}^2|$ (right) are also shown with an $1 \sigma$ band. The rate-only result for $\sin^2 2\theta_{13}$ is shown by the cross. The results from Daya Bay \cite{DB-spect} and Double Chooz \cite{DC-recent} are also shown for comparison.} 
\label{fig:contour-allowed}
\end{center}
\end{figure}

Figure \ref{fig:spectra-oscillation-fit} shows the background-subtracted, observed spectrum at far detector compared to the one expected for no oscillation and the one expected for the best-fit oscillation at the far detector. The expected spectra are obtained by weighting the spectrum at near detector with the oscillation or no oscillation assumptions using the measured values of $\theta_{13}$ and $|\Delta m_{ee}^2|$.
The observed spectrum shows a clear energy-dependent disappearance of reactor $\overline{\nu}_e$ consistent with neutrino oscillations. 
Figure \ref{fig:contour-allowed} shows 68.3, 95.5, and 99.7\% C.L. allowed regions for the neutrino oscillation parameters $|\Delta m_{ee}^2|$ and $\sin^2 2\theta_{13}$. 
The results from other reactor experiments \cite{DB-spect, DC-recent} are compared in the figure.

\begin{figure}[hbt]
\begin{center}
\includegraphics[width=0.47\textwidth]{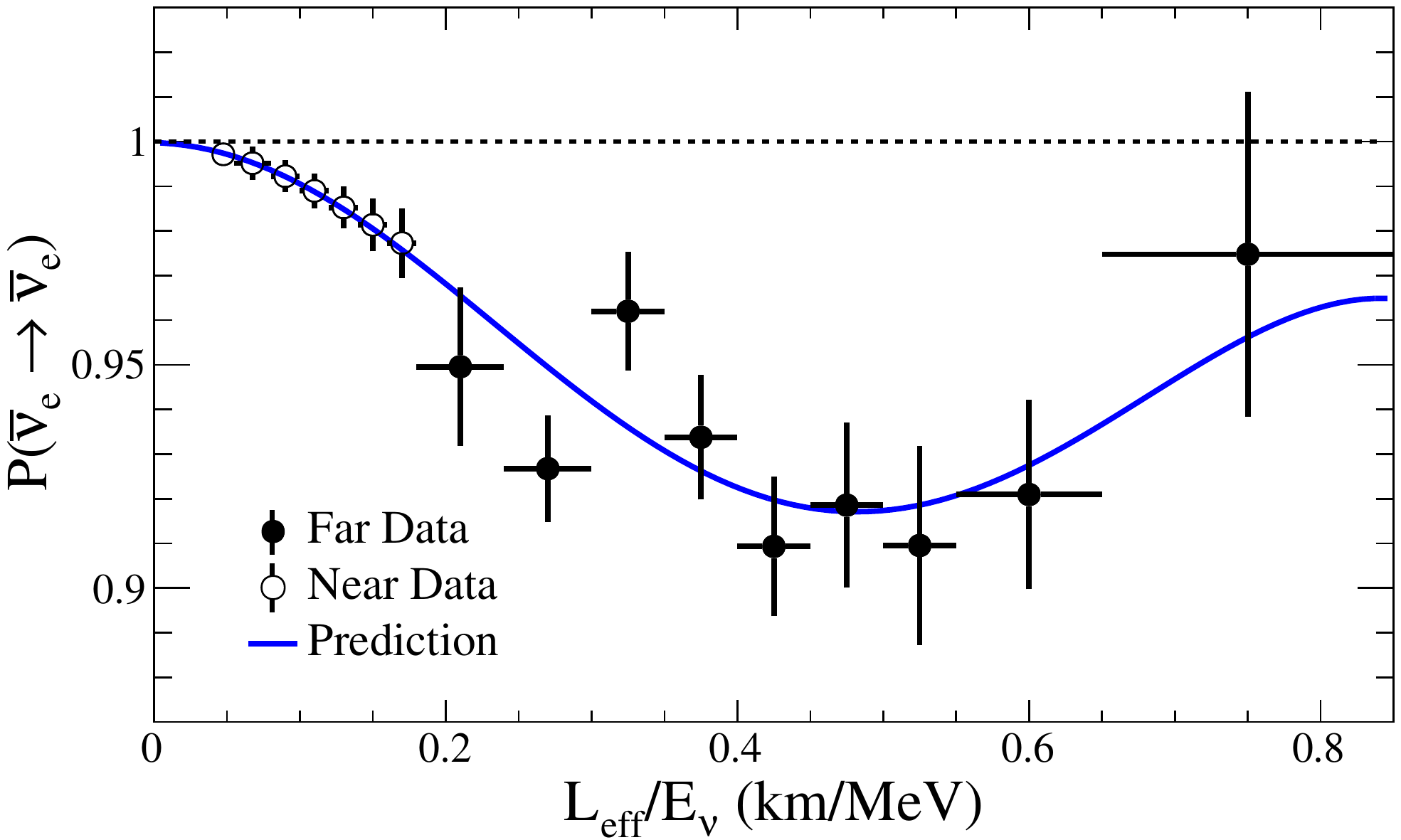}
\caption{Measured reactor $\overline{\nu}_e$ survival probability in the far detector as a function of $L_{\rm eff}/E_{\nu}$. The curve is a predicted survival probability, obtained from the observed probability in the near detector, for the best-fit values of $|\Delta m_{ee}^2|$ and $\sin^2 2\theta_{13}$. The $L_{\rm eff}/E_{\nu}$ value of each data point is given by the average of the counts in each bin.}
\label{fig:baseline-energy}
\end{center}
\end{figure}

Figure \ref{fig:baseline-energy} 
shows the measured survival probability of reactor $\overline{\nu}_e$ as a function of an effective baseline $L_{\rm eff}$ over $\overline{\nu}_e$ energy $E_{\nu}$ in the far detector, in a good agreement with the prediction that is obtained from the observed distribution in the near detector, for the best-fit oscillation values. This result demonstrates clear $L_{\rm eff}/E_{\nu}$-dependent disappearance of reactor $\overline{\nu}_e$, consistent with the periodic feature of neutrino oscillation. Note that $L_{\rm eff}$ is the reactor-detector distance weighted by the multiple reactor fluxes, and $E_{\nu}$ is converted from the IBD prompt energy. The measured survival probability is obtained by the ratio of the observed IBD counts to the expected counts assuming no oscillation in each bin of $L_{\rm eff}/E_{\nu}$.  

In summary, RENO has observed clear energy-dependent disappearance of reactor $\overline{\nu}_e$ using two identical detectors, and obtains $\sin^2 2\theta_{13} = 0.082 \pm 0.010 $ and $|\Delta m_{ee}^2| = (2.62^{+0.24}_{-0.26} )\times 10^{-3}$~eV$^2$ based on the measured periodic disappearance expected from neutrino oscillations. 
Several improvements in energy calibration and background estimation have been made to reduce the systematic error of $\sin^2 2\theta_{13}$ from 0.019 \cite{RENO} to 0.006. With the 500 day data sample together, RENO has produced a precise measurement of the mixing angle $\theta_{13}$. It would provide an important information on determination of the leptonic CP phase if combined with a result of an accelerator neutrino beam experiment \cite{T2K}.

The RENO experiment is supported by the National Research Foundation of Korea (NRF) grant No. 2009-0083526 funded by the Korea Ministry of Science, ICT \& Future Planning. Some of us have been supported by a fund from the BK21 of NRF.
We gratefully acknowledge the cooperation of the Hanbit Nuclear Power Site and the Korea Hydro \& Nuclear Power Co., Ltd. (KHNP). 
We thank KISTI for providing computing and network resources through GSDC, and all the technical and administrative people who greatly helped in making this experiment possible.

\end{document}